\documentclass[numreferences]{kluwer}

\usepackage{graphicx}              

\newcommand{\ket}[1]{\mbox{$|#1\protect\rangle$}}

\def\beq{\begin{equation}}
\def\eeq{\end{equation}}
\def\bea{\begin{eqnarray}}
\def\eea{\end{eqnarray}}

\def\ie{{\it i.e.}}
\def\eg{{\it e.g.}}

\begin{document}
\begin{opening}
\title{IS THERE MORE TO T?}
\subtitle{Why Time's Description in Modern Physics is Still Incomplete}
 
\author{A. C. ELITZUR}
\author{S. DOLEV}
\institute{The Program for History and Philosophy of Science\\
           Bar-Ilan University, 52900 Ramat-Gan, Israel}
\runningtitle{IS THERE MORE TO T?}
\runningauthor{ELITZUR and DOLEV}
\begin{abstract}
We present some novel results indicating that time's description in present-day physics is deficient. We use Hawking's information-erasure hypothesis to counter his own claim that time's arrow depends only on initial conditions. Next, we propose quantum mechanical experiments that yield inconsistent histories, suggesting that not only events but also entire histories might be governed by a more fundamental dynamics.
\end{abstract}
\end{opening}

\small

\section{The Stalemate}
\label{sec:time}
The problem associated with time's nature is well known. It stems from two aspects of time that cannot be reconciled: 

\begin{enumerate}
\item Time sharply differs from space in that in space, you can either move or stay put, and if you move you can do it in either direction. Not so with time. You cannot remain at the same moment, neither return to earlier moments. Time seems, then, to constantly move. 

\item The last sentence in nonsensical. Time cannot move neither can anything move in time, as the very notion of movement (passage, flow, etc.) entails time. Just ask ``what is the speed of time's movement?" and the absurdity of the statement will become apparent. You can, of course, assume another time parameter of a higher order, but that will necessitate a yet higher time dimension and so on {\it ad infinitum}. 
\end{enumerate}

Mainstream physics has chosen to deal with this problem by simply dismissing time's passage altogether as illusory. In the Minkowski 4-dimensional spacetime, all events -- past, present and future -- coexist along time just as all milestones coexist along a road. Only a few heretic theories sought to incorporate time's transitory aspect within a new theory (see \cite{Elitzur99b} for a brief review). Frustratingly, however, the debate never became genuinely scientific. All physical observations are equally consistent with a Block Universe (``all events coexist along time") and with the hypothesis of Becoming (``events are created anew one after another"). In the absence of a decisive experimental test, both views remain in the realm of philosophy. 

In this article we discuss some works of ours that provide clues about some more profound aspects of time. Because of space limits, our report is somewhat telegraphic, but references are given to the extended works, published in other journals with full mathematical details and relevant references.

\section{Whence Time's Arrow?}

This problem is also familiar. The basic laws of physics are time-symmetric, as well as all basic interactions. And yet, our macroscopic world is clearly time-asymmetric, in compliance with the Second Law of Thermodynamics. How can numerous time-symmetric interactions give rise together to one overall time-asymmetry? Two answers are available:

\begin{enumerate}
\item {\it It's all a matter of initial conditions.} The universe's initial conditions, \ie, the big bang, are highly ordered, allowing entropy only to increase. One could conceive of the opposite case, namely, a universe whose big bang is seemingly disordered, but with such a perfect correlation between all particles that entropy will eventually decrease. In fact, if one dismisses time's passage, then our universe is just such a universe. If the universe's beginning and end coexist along time, then one can read its history in either direction of time!

\item {\it Perhaps the basic interactions are not completely time symmetric after all.} Perhaps there is a very slight asymmetry in every basic interaction, unobservable by present-day means but underlying the macroscopic asymmetry that emerges from innumerable microscopic interactions.
\end{enumerate}

Clearly, (1) is the mainstream view as expressed by Hawking, while (2) is a hypothesis, endorsed only by Penrose and a very few authors (for the controversy, see \cite{Hawking96}). 

Intriguingly, Hawking himself, in respect to another issue, holds a famous heresy without noticing that it clashes with his conservative account of time's arrow. We refer to his claim that black hole evaporation involves a complete erasure of the information of all the objects that have earlier fallen into the black hole \cite{Hawking96}. We do not consider ourselves competent to voice an opinion in this debate. We only wish to point out that ``information-loss" is synonymous with ``indeterminism." But then, {\it for any closed system that contains an indeterministic event, that system's entropy can only increase, regardless of its initial conditions, in accordance with the time arrow of the rest of the universe.}

We have demonstrated this argument (\cite{Elitzur99b,Elitzur99c,Elitzur99}) by the means of a computer simulation of an entropy increasing process: On a billiard table, one ball is set to hit a group of ordered balls at rest, dispersing them around. After repeated collisions between the balls, the energy and momentum of the first ball is nearly equally divided between the balls. Although this state looks random, the correlations between the balls are strict: Reversing their momenta will time-reverse the scattering, resulting in a convergence back into the initial ordered state, ejecting back the first ball with the entire kinetic energy content of the system.

Obviously, this time-symmetry strictly necessitates perfect determinism. But what if determinism fails, say, if one ball's position is not a direct consequence of its previous state? We tried that by slightly disturbing the trajectory of one ball during the simulation. The entropy increasing process seemed to be the same, resulting in a similarly disordered state. However, when we applied the same disturbance to the time-reversed process, the return to the ordered initial state failed. 

In short: Entropy increasing processes do not require any special initial correlations, while entropy decreasing processes do -- they are extremely sensitive to any disturbance. This difference is so well-known to be a truism, but its straightforward bearing on time's arrow has not been noticed yet: {\it If physics ever proves that determinism does not always hold -- that some processes contain a truly random element -- it would follow that entropy always increases regardless of the system's initial conditions.} An intrinsic time-arrow must then emerge in any system, even in a closed one, independent of the initial conditions but congruent with the time arrow of the entire universe outside, from which that system is supposed to be shielded. If Hawking's information-loss conjecture turns out to be correct, its bearing on time's nature would be much more far-reaching than his otherwise-orthodox viewpoint allows. 

\section{Is Time symmetric at the Quantum Level?}
\label{sec:RPE}

\begin{figure}
\begin{center}
\includegraphics[scale=0.55]{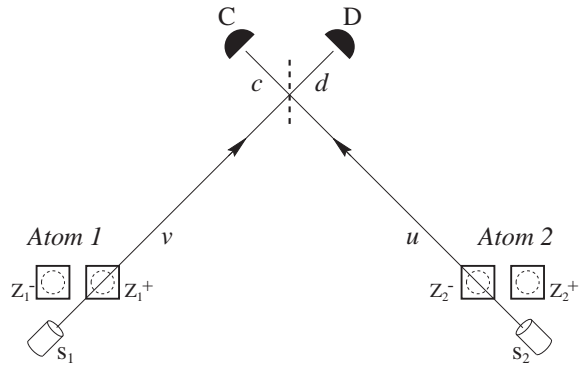}
\caption{Entangling two atoms by a future interaction (a single photon emitted by two sources) C and D denote constructive and destructive interference.}
\label{fig:RPE}
\end{center}
\end{figure}

Despite the apparent time-asymmetry associated with any measurement, the formalism of quantum theory is time symmetric. Recently, we have proposed an experiment that takes this time-symmetry to the extreme. 

The famous EPR experiment (\cite {EPR35}) involves two distant particles, emitted from the same atom. By spin conservation, they have opposite spins. On the other hand, by the formalism of QM, these spins are undecided until they are measured. Hence, measuring one of them must instantly determine the spin of the other, in apparent defiance of special relativity. Indeed, a celebrated theorem by Bell (\cite {Bell64}) proves that the correlations between the spins could not have been determined earlier than the very moment of measurement. 

Consider now another quantum effect, namely, the interference of light coming from different sources \cite{Paul86}. Though less famous, this effect is no less astonishing: When the radiation involved is sufficiently weak, then even the detection of a single particle could display interference pattern, as if one and the same particle ``has originated'' from two distant sources! 

Next, let us inbreed the two experiments (Figure \ref{fig:RPE}). Let two atoms be placed on the two possible routes of the two ``half photon"s. Let each atom be superposed in two boxes (\eg, by taking a spin 1/2 atom and splitting it according to it's $z$ spin). The boxes are transparent for the photon but opaque for the atoms. 

In 50\% of the cases, one of the atoms will ``choose" (or ``collapse") to reside in the box that intersects the path that the photon ``has chosen" too. These cases will result in scattering of the photon and will be discarded. In 25\% of the cases, however, one of the atoms will ``choose" to reside in the box that intersects one of the photon's possible paths, but the photon will ``choose" the other path. Here, the photon's interference will be disrupted, since one of its paths has been blocked by one of the atoms. On the other hand, the fact that the photon has arrived to the detector means that the other path was not blocked, \ie, that the other atom ``chose" not to intersect the other path. But {\it which} path? {\it Which} atom? Recall, now, that this is quantum mechanics, hence the ignorance about the atom is not merely epistemological but ontological. In other words, the very uncertainty about the positions of the atoms -- \ie, the question which atom lies in the intersecting box and which lies in the non-intersecting one -- suffices to physically entangle them in a full EPR state: 
\beq
\Psi = {1 \over \sqrt 2}(\ket{Z_1^-} \ket{Z_2^+} - \ket{ Z_1^+} \ket{ Z_2^-})
\eeq

Hence, tests of Bell's inequality performed on the two atoms indicate, just as the EPR, that the spin value of each atom depends on the choice of spin direction measured on the other atom, no matter how distant. The novelty in the present setting is that, unlike the ordinary EPR, where the two particles have interacted earlier, here the only common event lies in their {\it future}, namely, the detection of the single photon that might have visited either one of them. 

The heavy use of anthropomorphic and counterintuitive notions such as a photon ``choosing" to ``have originated" from a certain source might sound suspicious to readers who are accustomed to a more prudent language. The findings, however, are no less striking when described in strictly technical terms, as in our extended article \cite{Elitzur02}. 

This inversion of the EPR setting brings to mind some ingenious ``transactional" interpretations of QM (\eg, Aharonov \cite{Yakir90,Yakir95}, Cramer \cite{Cramer86}, Costa de Beauregard \cite{Costa87}), that proposed that each quantum interaction is the result of two interactions, one going forwards in time and the other going backwards, complementing one another so as to produce the observed quantum peculiarities. Thus, the EPR experiment is explained as an interaction extending along a spacetime zigzag between the two particles through the common past. It seems that our inverse EPR is particularly amenable to such an interpretation. It is the later detection of the photon that entangles the two atoms, even though their interactions with the photon have occurred earlier. Further variations of this experiment \cite{Elitzur02} also add to it Wheeler's ``delayed choice" (\cite{Wheeler78}) element, with the difference that, in our setting, the effect on the past leaves physically detectable traces. 

\section{Does the Wave Function Move Consequentially?}
\label{sec:atemporal}

Hardy \cite{Hardy92a,Hardy93} has proposed a very intriguing experiment in which a single photon traverses a Mach-Zender Interferometer (MZI) while an atom is superposed in two boxes, one of which is positioned across one of the photon's routes. If the photon ends up showing that the interference within the MZI was disrupted, it means that a) the atom must have been in the intersecting box in order to disrupt the interference, yet b) the photon must have taken the {\it other} route, otherwise it would have been scattered by the atom! The intriguing thing is that, although the photon has taken one path, the state of the atom positioned on the other path has been changed in a physically measurable way, although a very subtle one: It has lost its initial superposition and assumed a definite state.

\begin{figure}
\centering
\includegraphics[scale=0.55]{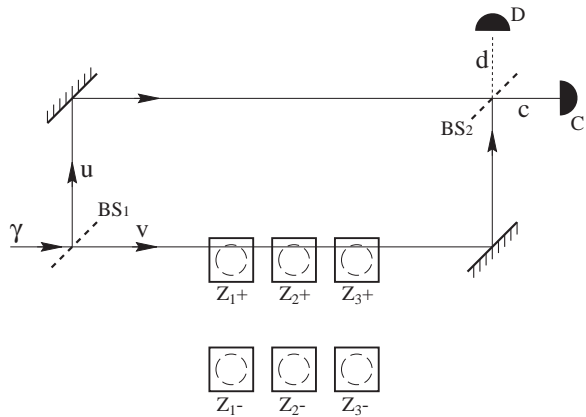}
\caption{Non sequential interaction: i) the photon ends up in D, indicating that one of the superposed atoms is in the intersecting box, ii) the middle atom is found to be in the intersecting box, iii) all the other atoms return to their superposition (\eg, manifesting full interference), as if noting has ever interacted with them.}
\label{fig:atemp}
\end{figure}

Hardy argued that this experiment provides support for the ``guide wave" interpretation of quantum mechanics. It is the half wave plus particles, he reasoned, that went through the left arm, while the other empty half wave went through the right arm and ``collapsed" the atom. 

This argument challenged us \cite{Dolev00}  to modify the device in a way that may enable one to empirically distinguish between the ``guide wave" interpretation and the ordinary ``collapse" interpretation. In the latter interpretation, the wave function is evenly split, then goes through both MZI arms, and then, upon encountering a measuring object, vanishes from one arm and becomes fully materialized in the other. In order to test both interpretations against each other, we have simply replaced the one atom in Hardy's version with a few (say, three) atoms (as in Figure \ref{fig:atemp}). This, we hoped, will enable us to trace the photon's subtle action along space. True, the guide wave interpretation is often formulated in a way that yields the same experimental predictions as ordinary quantum theory, but we still felt that the results might strain one of the two ontologies so as to favor the other.

Much to our surprise, the results supported neither ``guide wave" not ``collapse," but demonstrated an even more intriguing effect. As in Hardy's version, here too, if the photon indicates that interference was disrupted, then, with 100\% certainty, {\it one} of the atoms has ``collapsed" into the intersecting box. However, it can be {\it any} of the N atoms, not necessarily the first. Worse, once we have measured one of the atoms and found it in the intersecting box, all the other atoms return to their original, undisrupted, superposition state. Consequently, if we do not measure these atoms' positions but reunite the boxes and perform an ``interference" measurement, the atoms will {\it always} exhibit full interference, as if no photon has ever interacted with them!  

This result severely offends ordinary spatio-temporal notions. If one assumes that the photon's wave function has interacted with the particular atom we've measured so as to ruin its interference, how come that all the other atoms in the row, positioned before and after that particular atom, seem to have never been affected?

\section{Does Measurement Always Give Consistent Histories?}
\label{sec:inconsistent}

Another offence to the ordinary temporal notions comes from our above inverse EPR experiment (section \ref{sec:RPE}). At first sight, that setup seems to add support to the claim that quantum mechanical interactions are transactions between earlier and later events, thereby lending support to a static view of spacetime (see section \ref{sec:time}). However, a variation of that experiment gives a result that hardly accords with the conventional description of spacetime.

Let us first recall the essence of Bell's nonlocality proof (\cite {Bell64}) for the EPR experiment. Consider three spin directions, $x$, $y$, and $z$. On each pair of EPR particles, one out of these directions should be measured at random on each particle. Let many pairs be measured this way, such that all possible combinations of $x$, $y$, and $z$ are performed. Then let the incidence of correlations and anti-correlations be counted. If quantum mechanics is correct, all same-spin pairs will yield correlations, while all different-spin pairs will yield 50\% correlations. Indeed, this is the result obtained by numerous experiments to this day. By Bell's proof, such a result could not have been pre-established. Hence, an instantaneous influence between the particles must take place at the moment of measurement. 

Let us now apply this method to our inverted EPR. Each atom's position, namely, whether it resides in one box or the other, constitutes a spin measurement in the $z$ directions (as it has been split according to its spin in this direction). To perform the $z$ measurement, then, one has to simply open the two boxes and check where the atom is. To perform $x$ and $y$ spin measurements, one has to re-unite the two boxes under the inverse magnetic field, and then measure the atom's spin in the desired direction. Having randomly performed all nine possible pairs of measurements on the pairs, and using Bell's theorem, one can prove that the two atoms affect one another instantaneously, with the difference that they share an event not in the past, as in the ordinary EPR, but in the future. 

However, a very bizarre situation now emerges. In 44\% (\eg, $4 \over 9$) of the cases (assuming random choice of measurement directions), one of the atoms will be subjected to $z$ measurement -- namely, checking in which box it resides -- while the other atom will be subjected to $x$ or $y$ -- namely, reuniting its two boxes and then measuring another spin direction. Suppose, then, that the first atom was found in the intersecting box. This means that {\it no photon has ever crossed that path}. But then, by Bell's proof, the other atom is still affected nonlocally by the measurement of the first atom. But then again, if no photon has interacted with the first atom (remember that we post-selected out all cases of scattering), the two atoms share no causal connection, in either past of future!

Like the wave function's inconsequential behavior in section \ref{sec:atemporal}, this experiment yields a history that is not consistent: One atom indicates that the photon has taken only one path, while the other atom's state proves that both atoms have been visited by the same photon.

\section{A Speculation}

What alternative picture of time might eventually emerge from these cracks in the prevailing paradigm? Fully aware of this question's pretentiousness, we nevertheless risk a speculation. 

First, concerning Hawking's information-loss conjecture, we reiterate that, if this conjecture turns out to be correct, its bearing on the present picture of time would be devastating. Time's asymmetry would turn out to be inherent to all physical processes, rather than an artifact of boundary conditions. Perhaps this conclusion might be less surprising for anybody who keeps in mind the still unexplained CP violation exhibited by neutral kaons, which, by CPT invariance, entails a fundamental violation of T. Consequently, if a subtle time-asymmetry is inherent to physical interactions themselves, the orthodox picture of time as a mere dimension looses much of its conviction. 

Our next comment concerns the apparently inconsistent histories implied by the experiments in sections \ref{sec:atemporal} and \ref{sec:inconsistent}. Earlier (section \ref{sec:RPE}) we have mentioned the ``transactional" interpretations (\cite{Cramer86,Costa87}, that, by invoking retarded-plus-advanced actions, offer a simple and elegant explanation for many spatial and temporal peculiarities manifested by QM. However, these interpretations adhere to the ``Block Universe" view and deny any dynamics to spacetime itself. Is it possible to have a transactional model that allows some dynamics to occur in spacetime itself?

We envisage such a model, although at present it is still highly tentative. From general relativity, we take the concept of spacetime as a real physical thing, namely, a four-dimensional manifold of world lines with their corresponding spacetime curvatures. Within this geometric picture, the transactional interpretations fit in very naturally. Where we break new ground is in proposing that this spacetime is not static. Perhaps it, too, is subject to some subtle dynamics, that is changes affect not only events but also {\it entire histories}. Then, time's asymmetry will be anchored in that dynamics governing spacetime itself (\eg, the alleged progress of the ``Now''). Also, quantum mechanical experiments yielding apparently inconsistent histories, as those described above, would give rise to an account like ``first a retarded interaction brings about history $t_1x_1, t_2x_2,...$ and then an advanced interaction transforms this history into $t_1x'_1, t_2x'_2,...$." Such a model will be better capable of explaining quantum peculiarities of the kind described above, as well as a few other surprising results discovered lately by similar techniques (\cite{Yakir90,Yakir95}). But then, such a model will be nothing short of a new theory of spacetime.

\bibliographystyle{unsrt}

\bibliography{NATO-QUANT-PH}

\end{document}